\documentstyle[12pt,aasms4]{article}
\received{1997.08.20}

\begin{document}

\title{\bf EMERGENCE OF A TWISTED MAGNETIC FLUX BUNDLE
AS A SOURCE OF STRONG FLARE ACTIVITY}
\author{\bf TAKAKO T. ISHII$^{1,2}$ , HIROKI KUROKAWA$^{1}$ 
AND \\ TSUTOMU T. TAKEUCHI$^{2,3}$}
\affil
{ 
$^1$ Kwasan and Hida Observatories, Kyoto University, Yamashina-ku, 
Kyoto 607, JAPAN \\
$^2$ Department of Astronomy, Kyoto University, Sakyo-ku, 
Kyoto 606-01, JAPAN \\
$^3$ Research Fellow of the Japan Society for the 
Promotion of Science
}
\centerline{E-mail: ishii@kusastro.kyoto-u.ac.jp}
\centerline{REVISED November~~28th~~1997}

\begin{abstract}

Sunspot proper motions and flares of a super active region 
NOAA 5395, which was the biggest and the most 
flare-active region in the 22nd sunspot cycle,
were analyzed in details.
We measured sunspot proper motions by using the H$\alpha - 5.0$ \AA 
$\;$ images obtained with the 60-cm Domeless Solar Telescope (DST) 
at Hida Observatory, Kyoto University and found some peculiar 
vortex-like motions of small satellite spots successively 
emerged from the leading edge of this sunspot group.
To explain these motions of small sunspots, 
we proposed a schematic model of the successive emergence of 
twisted and winding magnetic flux loops coiling around a trunk 
of magnetic flux tube.
The location of the strongest flare activity was found to coincide 
with very the site of the vortex-like motions of sunspots.
We conclude that the flare-productive magnetic shear is produced by 
the emergence of the twisted magnetic flux bundle.
Magnetic energy is stored in the twisted flux bundle
which is originally formed in the convection zone and 
released as flares in the course of the emergence of the twisted 
flux bundle above the photosphere.

\end{abstract}


\section{INTRODUCTION}

Characteristics of flare-productive sunspot groups have been 
studied by many authors 
(Zirin \& Tanaka 1973; Hagyard et al. 1984; Kurokawa 1987; 
Zirin \& Liggett 1987; Tanaka 1991; and Kurokawa 1991, 1996).
In these studies, the authors pointed out that
sheared configuration of magnetic field and 
$\delta$-type configuration of sunspot group are 
essential characteristics for strong flare activities.

From statistical studies, however, Shi \& Wang (1994) found 
that 23 \% of $\delta$-type sunspot groups produced X-class 
flares, and Chen et al. (1994) found that even strong 
magnetic shear of more than 80-degree shear angle is not a 
sufficient condition to produce flares.

On the other hand, Tanaka (1991) suggested that an 
emergence of a twisted magnetic flux rope produces strong flares.
Kurokawa (1987, 1991, 1996) also concluded that strong flare 
activities occur whenever a magnetic shear is rapidly 
developed by a successive emergence of twisted magnetic loops.
Zirin \& Liggett (1987) showed that big flares occurred in 
$\delta$-groups which emerged as continents.

These studies suggest that flare-productivity of a sunspot 
group depends on formation process of the $\delta$-type 
configuration or the magnetic shear.
This processes can be examined by measuring of sunspot 
proper motions.

A large active region observed in March 1989 (NOAA 5395) 
was a highly flare-productive sunspot group with 
$\delta$-type configuration.
The region produced about two hundred flares including ten 
X-class flares.
All X-class flares were observed in three particular regions.
Wang et al. (1991) studied this region using 
Magnetograms, Dopplergrams and monochromatic images 
mainly obtained at Big Bear Solar Observatory (BBSO) and 
Zhang (1995) also studied using Magnetograms obtained 
at Huairou Solar Observing Station (HSO).

We measured proper motions of sunspots in NOAA 5395 by using 
H$\alpha$ images obtained with the 60-cm Domeless Solar 
Telescope (DST) at Hida Observatory, Kyoto University 
in order to study the formation process of $\delta$-type 
configuration and magnetic shear in the region.
In this study we demonstrate that twisted magnetic flux tubes 
successively emerged at the preceding edge of the great sunspot 
group and that they played an essential role in the production 
of strong flare activities of NOAA 5395.

We describe the procedure of data analysis in Sec. 2 and show 
the obtained results of sunspot proper motions in Sec. 3.
In Sec. 4 we discuss the cause of peculiar sunspot motions 
found at the leading edge of the active region and propose a 
morphological model of an emerging twisted magnetic flux 
bundle which is essential to the strong flare activity.
Section 5 is devoted for our conclusion.

\section{DATA ANALYSIS}

\subsection{Observational Data}

The active region NOAA 5395 turned into the visible hemisphere 
on March 6 and turned off behind the western limb on March 19.
During this period, except on March 7, 13 and 17, we continuously 
observed the evolution of the region with DST.
Many H$\alpha$ monochromatic images were sequentially obtained 
with the Zeiss Lyot filter of 0.25 \AA $\;$ passband and a Nikon 
motor-drive film camera of DST in 10 wavelength ({\sl i.e.} 
H$\alpha \pm 0.0$ \AA, $\pm 0.3$ \AA, $\pm 0.5$ \AA, 
$\pm 0.8$ \AA, $\pm 1.2$ \AA $\;$ and $-5.0$ \AA).
In this study we mainly used H$\alpha -5.0$ \AA $\;$ images 
to measure sunspot proper motions. 

We digitized the film densities of the H$\alpha -5.0$ \AA $\;$ 
images by a film scanner and analyzed them by using IDL 
(Interactive Data Language) software on a personal computer and 
UNIX.
Times and numbers of measured images are listed in Table 1.
One example of the H$\alpha -5.0$ \AA $\;$ images is shown 
in Fig. 1.

We carefully identified individual sunspot umbrae of the group 
in successive images and gave numbers to them as shown in Fig. 1b, 
where the signs F and f stand for the {\sl following} magnetic 
polarity, P and p, the {\sl preceding} polarity 
and the sign '' : '' means uncertainness.
We used the magnetograph data supplied by Okayama Astrophysical 
Observatory (OAO), BBSO and HSO for the determination 
of magnetic polarities of the sunspots.

\subsection{Calculation of Latitude and Carrington Longitude of 
Sunspot Umbrae}

The field of view of the H$\alpha$ filtergraph of DST is a circle 
of about 390 $''$ diameter.
The heliocentric coordinate of the center of the circle is recorded 
for each observation.

Referring to the center of the circle, we determined the heliocentric 
coordinate of individual sunspot umbra in IDL pictures.
The umbra position was defined as the  center of gravity position 
of the umbra, which is determined by drawing the density contour 
of each umbra in an IDL picture.
By means of coordinate transformation we finally obtained 
heliographic latitude and Carrington longitude of each umbra.

During this observation the pointing accuracy of DST was 
about 10 $''$, and this limits the accuracy of absolute positions 
of umbrae. 
Then we defined the  center of gravity of a triangle made by 
three umbrae F1, F3 and P1 as a reference point and 
determined the relative positions of all sunspot umbrae 
referred to this point.
The mean heliographic latitude of the reference point averaged 
over all measured images is $33.^\circ 6$ N and its standard 
deviation is 1.0 degree.
By excluding the images whose deviation exceeds $1 \sigma$, 
we get $33.^\circ5$ N as the average latitude of the reference 
point.

The longitudinal positions are significantly 
influenced by the differential rotation.

Tang(1981) gave the differential rotation rate for the 
sunspot groups at high ($28^\circ \sim 40^\circ$) latitude
by the formula
\begin{equation}
\Omega = 14.37 -2.60 \sin^2 \theta \quad{\rm deg/day},
\end{equation}
where $\Omega$ is the sidereal rotation rate.
We would like to know the synodic rotation rate.
According to Zirin (1989), the synodic rotation rate ($\omega$)
and the sidereal rate ($\Omega$) differ by the 
Earth's orbital velocity of 0.9865 deg/day; 
\begin{equation}
\omega = \Omega - 0.9865 \quad {\rm deg/day}.
\end{equation}
And we find that the synodic rotation rate at the latitude of 
$33.^\circ 5$ N is 12.59 deg/day.

On the other hand, the heliographic longitude of the disk 
center ($L_0$) decreased by the rate of 13.18 deg/day.
Considering the difference of above two values
(i.e. 12.59 deg/day and 13.18 deg/day), 
the Carrington longitude of the reference point should 
change by the rate of 0.59 deg/day due to the 
differential rotation.

We first determined the heliographic latitude and 
Carrington longitude of the reference point for the image 
of 00:00 UT of Mrach 11, which are $33.^\circ 5$ N (latitude) 
and $254.^\circ 5$ L (longitude) respectively.
Then for all other images we calculated the Carrington longitude 
of the reference point by correcting the value of 
0.59 deg/day.
For each sunspot umbra, we adapted a different correction rate 
for the differential rotation corresponding to its latitude.
After these corrections for the differential rotation, we 
calculated velocities of proper motions of sunspot umbrae.
The term of umbra velocity used hereafter in this paper 
means a velocity of proper motion.

\section{RESULTS}

We noticed that the motion of F1 is quite different from those of 
other big umbrae (F3, F4 and P1); 
the umbra F1 moved southward or southwestward, while the others 
moved eastward or northwestward.
The umbrae F3, F4 and P1 moved toward east at velocities of 
$0.03 \; \sim \; 0.08$ km/s.
The umbra F1 moved southwestward at a velocity of 
about 0.05 km/s and it suddenly accelerated 
after 12 March.
The velocity of F1 reached at 0.11 km/s on March 14.
The other three big umbrae moved eastward at velocities of 
$0.03 \; \sim \; 0.08$ km/s.
These velocities are approximately the same as those of 
previous works, in which sunspot proper motions were 
measured in various active regions 
(van Driel-Gesztelyi \& Petrovay 1990, 
van Driel-Gesztelyi {\sl et al.} 1993, 
van Driel-Gesztelyi {\sl et al.} 1994, 
Schmieder {\sl et al.} 1994, and 
Herdiwijaya, Makita, \& Anwar 1997).

According to Herdiwijaya et al. (1997), 
the longitudinal velocities of sunspots varies with 
their Z\"{u}rich classes; 
the average velocity of GHJ classes (i.e. decaying phase) 
is smaller than that of AB classes (i.e. emerging phase). 
The {\sl following} polarity sunspots of Z\"{u}rich AB classes 
show the average longitudinal 
velocity of 88 m/s.
On the other hand, the average longitudinal velocities 
of EF classes (i.e. well-developed phase) and GHJ classes are 
18 m/s and 21 m/s respectively.
The average longitudinal velocity of {\sl following} polarity spot F1 
is 57 m/s for the period from 9 to 15 March.
This means that the magnetic flux tube forming sunspot F1 is 
still emerging from the subphotospheric layer 
in our observation period.

We also noticed conspicuous motions of small sunspot umbrae around 
the leading sunspot F1.
They successively appeared at the south and southeast edge of 
F1 and moved eastward.
We examined their motions in details. 
By calculating the displacements of these small umbrae 
between successive frames of images in each day, 
we determined their velocities of proper motions. 
We also calculated their velocities relative to F1 
(relative velocities, hereafter).
The results obtained only for images of good seeing are 
presented in Fig. 2, where
arrows indicate the relative velocities of the small umbrae.

On March 9 (Fig. 2a), small umbrae near F1 
(p1 and  p2) moved southward and 
others (p3, p4 and p5) moved eastward.
The relative velocities of these umbrae were all 0.15 km/s.
On March 10 (Fig. 2b), small umbrae (p:6, f:1) near F1 
moved southward at relative velocities of 
$0.15 \; {\rm and} \; 0.30$ km/s, respectively.
A $p$-polarity  umbra (p7) moved southeastward 
at a relative velocity of $0.10$ km/s. 
A $p$-polarity  umbra (p9), which moved toward P1, 
had a relative velocity of 0.20 km/s.
A $p$-polarity umbra (p8) moved eastward 
at a relative velocity of 0.10 km/s.
Small umbrae at the north side of F1 moved eastward along 
the channel between F1 and F2.
On March 11 (Fig. 2c), small umbrae at the east and 
the southeast sides of F1 moved eastward or northeastward 
at relative velocities of $0.30 \; \sim \; 0.40$ km/s.
Small umbrae at the north side and at the west side of F1 
($f$-polarities) also moved eastward or northeastward.
On March 12 (Fig. 2d), small umbrae at the east and 
southeast side of F1 also moved eastward at much higher 
relative velocities of $0.30 \; \sim \; 0.60$ km/s.
Small umbrae ($f$-polarities) at the north side 
and at the northwest side of F1 
also moved east or northeast along the north edge of F1.

We also calculated the relative velocities of 
proper motions of the small umbrae by using all available images 
for each day.
The average velocities obtained in each day are given in Fig. 3.
It is found that the two results of Fig. 2 and Fig. 3 agree with 
each other.
From these results we summarize main characteristics of 
proper motions of small umbrae as follows.

At the south and southeast edges of F1, small umbrae successively 
emerged and moved toward P1 through the east side of F1. 
Appearing near F1, they first moved southward, 
turned to the east and finally moved northeastward to approach P1.
It means they moved along curved trajectory like 
an outflowing vortex.
Such emergences and motions of small sunspots 
continued with their speed accelerated 
from 9 through 12 of March.
Spot P1 became bigger in its umbra area
from 10 through 12, March.
Wang et al. (1991) also found that $p$-polarity umbrae 
at the east side of F1 merged into P1.
We found that spot P2 newly formed on 14 Mar at 
the region where small emerging sunspots or emerging magnetic 
flux successively converged from 12 through 14 March (see Fig. 1b).

Some small umbrae emerging at the south west edge of F1 
(i.e. f6, f10, f11, f13, f14, and f15)
first moved northward at the west side of F1 and moved 
eastward or northeastward along the north edge of F1 
from Mar 10 through 12.
They converged into a few larger umbrae at the northeast 
side of F1 on March 12.
By March 14 the northeast part of F1 has been separated from 
its leading part and moved northeastward to form F5 umbra.

\section{DISCUSSIONS} 

\subsection{A Schematic Model of Emerging Twisted Flux Tubes}

The most remarkable result found in the previous section is the 
vortex-like motions of small sunspot umbrae around the leading 
sunspot F1.
Pairs of small umbrae of different magnetic polarities 
successively emerged 
at the leading edge of F1 (i.e. at the south 
or southwest edges of F1) from 9 through 12, March.
These umbrae moved clockwise toward the southwest of P1, 
where growth of P1 was observed from 10 through 12, March 
and P2 was formed on March 14.
At the west and north sides of F1, small umbrae moved 
anticlockwise and formed F5 by merging with some parts of decaying F1.
These motions are schematically summarized in Fig. 4a.
At the east side of F1, small umbrae, most 
of which are  $p$-polarities, moved clockwise and formed P2.
At the west and north sides of F1, small umbrae, all $f$-polarities, 
moved anticlockwise and formed F5 by merging with some parts 
of decaying F1.
Notice vortex-like lines and structures of small 
sunspot umbrae and penumbrae surrounding F1 spot in Fig. 1a.
Wang et al. (1991) also found the same 
vortex-like motions of magnetic features in this region 
using Magnetograph data.

For the explanation of these peculiar vortex-like motions, 
we propose a schematic model of emerging twisted flux tubes 
given in Fig. 4b.
The model is characterized by the successive emergence of 
twisted flux tubes coiling around the main trunk of 
flux tube F1.
The bundle of flux tubes is twisted and each flux tube 
is also twisted.
As Parker (1955) pointed out, the magnetic buoyancy 
raises up the system of twisted magnetic-flux bundle.

The successive emergence of the inclined and twisted flux bundle
can explain the vortex-like motions of small sunspots 
both at the southeast and the northwest sides of F1 
as shown in Fig. 4a.
In Fig. 4b planes $P_{t_1} {\rm and} P_{t_2}$ show the 
positions of the photosphere relative to the 
emerging flux tubes for the different days.
The cross sections of flux tubes on each photospheric plane 
correspond to sunspots.
As the coiling loops successively emerge, the small 
$p$-polarity and $f$-polarity sunspots can be seen to move 
clockwise and anticlockwise, 
at the southeast side and at the northwest side of F1 respectively.
Notice that small $p$-polarity sunspots seen on the planes $P_{t_1}$ 
are all converged into the sunspots P2 on $P_{t_2}$ 
and that $f$-polarity ones are converged into the sunspot F5.  
These are very consistent with the observations 
summarized at the  beginning of this section.

In addition, it is interesting to notice another evolutional
feature of F1.
As studied in section 3, the southwestward motion 
of the umbra F1 suddenly accelerated after 12 March.
We observed the drastic change of the shape of F1
at the same time (see Fig 1b); 
F1 suddenly elongated and started to decay after 12 March.
We also observed the rapid growth of P2 and F5 umbrae 
between 12 and 14 March.
This time coincidence between the decay of F1 and the 
growth of P2 and F5 strongly suggests a close causal relation 
between them : 
The umbrae F1 was stable until 12 March because it was bound 
by the coiling flux tubes.
When the coiling flux tubes already emerged above the 
photosphere after 12 March, however, the trunk of 
flux tube F1 suddenly expanded, or elongated and 
started to decay.

Furthermore, it must be worthwhile here to mention about
a differential emergence of magnetic loops. 
Suppose the gas pressure and magnetic field strength
in the tube are $p_{\rm i}$ and $B_{\rm i}$ and the surrounding 
external pressure at the same height is $p_{\rm e}$.
Then the lateral total pressure balance implies
\begin{equation}
p_{\rm e} = p_{\rm i} + \frac{B_{\rm i}^2}{8 \pi}.
\end{equation}
If the temperature ($T$) is uniform and the corresponding
densities are $\rho_{\rm i}, \rho_{\rm e}$, Equation (3) becomes
\begin{equation}
\frac{k_{\rm B} T \rho_{\rm e}}{m} = \frac{k_{\rm B} T \rho_{\rm i}}{m} + 
\frac{B_{\rm i}^2}{8 \pi},
\end{equation}
where we take the perfect gas law ;
$p=k_{\rm B} \rho T/m$ ($k_{\rm B}$ is the Boltzmann constant and $m$ is 
the mean particle mass).
The plasma in the tube feels a resultant buoyancy force 
of $(\rho_{\rm e} - \rho_{\rm i})g$ per unit volume, which tends 
to make the tube rise.

The trunk and coiling loops in Fig 4b are considered to have different 
magnetic pressures ($B_{\rm i}$) and different densities 
($\rho_{\rm i}$).
Different densities give different buoyancy force, so the system
rises differentially.
Difference of the shapes of flux tubes (i.e. different 
curvatures of flux tubes) may also make the system rise
differentially.

In spite of this consideration, the schematic drawing like 
Fig. 4b may give the impression that the whole system 
emerges at the same speed, because we do not draw how 
differentially the loops rise.
We need more observational data of better temporal- and spatial-
resolutions before we can discuss more realistic 
structures of successively emerging flux loops.
And this is an important subject to be attacked in future works.
Figure 4b, however, well shows the essential characteristics of our model,
i.e. an emergence of a twisted magnetic flux bundle.
It does not influence our conclusion how differentially the flux loops rise.

\subsection{Flare Activity in NOAA 5395}
 
According to Solar Geophysical Data (SGD), 
about two hundred flares occurred in this active region.
Some example of flares observed with DST at Hida 
Observatory are shown in Fig. 5.
The positions of the flares given in SGD were transformed 
to our coordinate system (heliographic latitudes and 
Carrington longitude) and 
these of the flares stronger than C-class in 
soft X-ray importance were plotted in Fig. 6, where 
the positions of the flares are given for each day 
(from 15:00 UT of the previous day 
to 15:00 UT of the day).
The relative velocities of the small sunspot umbra around F1 
are also shown in Fig. 6.
Notice that almost all flares occurred around F1 from 
9 through 12 March.
On 15 March, before which the twisted flux bundle has already 
emerged out, however, no strong flare occurred around 
F1 (Fig. 6).
On the contrary, some strong flares occurred along the 
sheared neutral line formed between P2 and F5 on 14 and 
15 March.

Especially noticeable in these two figures (Fig. 5 and 6) 
is that from 9 through 12, March, 
many M-class and C-class flares occurred at the leading edge 
and at the east of F1 where the vortex-like motions of the small 
umbrae were found.
The north side of F1 is also flare-productive especially on 12 March.
These results strongly suggest that 
the vortex motions of small spots around F1 or the 
successive emergence of twisted magnetic flux tubes 
should be sources of the strong flare activities in this region.

\subsection{Flare Energy Build-up Process}

The vortex-like motions of small satellite spots around F1
were explained with emergences of the twisted magnetic flux 
bundle(Sec. 4.1), 
and strong flare activities preferentially occurred around this 
emerging flux region (Sec. 4.2).
These evidence strongly suggest the existence of causal 
relations between the emergences of twisted magnetic flux tubes 
and the strong flare activities.
With successive emergence of twisted flux tubes, very complex 
magnetic field structures, which consist of 
many current loops, were formed around F1 and at the 
south of P1.
Notice the twisted dark loops at the east side of F1 
in the H$\alpha$ images of Fig. 7.
They have different orientations with each other, 
and they are considered to reconnect with each other 
and sometimes to produce flares as shown in Fig. 5.

In the convection zone, magnetic flux tubes are 
intensified by a dynamo action and 
twisted by convection in a rotating fluid.
When the twisted flux tubes emerge to the photosphere, 
the twists should get loose as shown in  Fig. 4 with 
decrement of the gas pressure.
Since the twisted magnetic tubes store much energy, 
the energy could be released in the violent flare activities 
in the F1 region of NOAA 5395.

The idea that twisted magnetic flux tubes store the energy 
for flares is first proposed by Piddington (1974), and 
several observational studies developing this idea have been 
published by Tanaka (1991), Kurokawa (1987, 1991, 1996), 
Wang (1994) and Wang et al. (1996).
Our study presents another clear evidence of the emergence 
of a strongly twisted and flare-productive magnetic flux bundle.
Especially important findings in this study are the outflowing 
vortex-like motions of small sunspots and extremely hot flare 
activity around the leading sunspot F1.
Such conspicuous motions of sunspots and associated 
hot flare activities have been hardly reported before, 
though Kurokawa et al. (1987) noticed a conspicuous 
growth of the leading 
part of $\delta$-type sunspot just before a great flare 
of April 25, 1984 in NOAA 4474.
Leka et al. (1994, 1996) also reported some vortex-like 
motions of small sunspots in an emerging flux region, but 
the region is not so active in flare production.
More examinations of the relation between the vortex motions of 
sunspot structures and flare activity are necessary in future 
observations and measurements of existing data.

\section{CONCLUSION}

Sunspot proper motions and flares of a super active region 
NOAA 5395, which is one of the biggest and the most flare-active 
regions in the 22nd sunspot cycle, were analyzed in details.
We measured the sunspot proper motions 
by using the H$\alpha$ images obtained with DST and 
found some peculiar vortex-like motions 
of small satellite spots successively emerging from the 
leading sunspot F1.
For the explanation of these motions of small sunspots, 
we proposed a schematic model of the successive emergence of 
twisted and winding magnetic flux loops coiling around 
a trunk of magnetic flux tube (Fig 4).

We also found most of flares preferentially occurred in this 
emerging  region around F1.
These results are consistent with our previous conclusion 
that the flare productive magnetic shear is produced by the 
emergence of twisted magnetic flux tubes (Kurokawa 1987, 1991, 
1996).
This study strongly suggests that the magnetic energy 
of a flare-productive sunspot group is stored 
in a twisted flux bundle which is originally 
formed in the convection zone.
The energy is released as flares in the course of 
the emergence of the twisted flux bundle above the photosphere.

Studies of  the typical flux tube geometry which has enough energy 
for big flares should be useful for forecasting flare occurrence.
It is still unclear, however, which type of emergence of 
twisted flux tube always produces strong flare activity.
We need more observational studies of the process of 
magnetic shear developments in more active regions by 
using proper motions of sunspots and evolutional changes 
of H$\alpha$ fine structures and photospheric vector 
magnetic field.

\acknowledgments

We would like to thank the anonymous referee for the useful comments 
which improved our paper.
We also thank Drs. T. Sakurai, F. Tang and 
H. Zhang for their providing magnetograms.
We are grateful to Drs. R. Kitai and Y. Funakoshi for 
their helps in the observations at Hida Observatory.
Dr. H. Hirashita contributed helpful discussions and advice.
One of the authors (TTT) acknowledges the Research Fellow 
of the Japan Society for Promotion of Science for 
Young Scientists.


\pagebreak

Fig. 1.--- 
The H$\alpha -5.0$ \AA $\;$ image obtained with DST 
at 23:19 on March 11. 
Notice the vortex-like lines of small umbrae and penumbra
structures surrounding a large sunspot umbra F1.
(b) Daily evolutional changes of NOAA 5395.
North is up, west is to right.
An interval of two hair lines corresponds to 100 $''$.
Sunspots measured their proper motions are given numbers.
the signs F and f stand for the {\sl following} polarity 
and the P and p, the {\sl preceding} polarity.
the sign '' : '' means uncertainness.
This region was elongated toward southwest and northeast.

Fig. 2.--- Velocities of small umbrae relative to F1 
measured for two images of good seeing.
An arrow in the upper right corner of each panel 
indicates the velocity scale of 100 m/s.
Vortex-like motions of small satellite spots are found.
(a) From 23:31 UT on March 8 to 05:10 UT on March 9. 
(b) From 00:37 UT to 05:46 UT on March 10. 
(c) From 23:24 UT on March 10 to 01:28 UT on March 11. 
(d) From 23:18 UT on March 11 to 02:48 UT on March 12. 

Fig. 3.--- The same as Fig. 2 but for all images on a day.

Fig. 4.--- 
(a) Schematic summary of observed vortex-like motions of 
small umbrae surrounding F1. 
Pairs of small umbrae successively emerged at the leading 
edge of F1.
At the east side of F1, small umbrae, mainly $p$-polarities, 
moved clockwise and formed P2.
At the west and north sides of F1, small umbrae, all $f$-polarities, 
moved anticlockwise and formed F5 by merging 
with some parts of decaying F1.
(b) A schematic model of emerging flux tubes.
The bundle of flux tubes is twisted and each flux tube is also twisted.
Planes $P_{t_1}$ and $P_{t_2}$ show the positions of the photosphere 
relative to the emerging flux tubes for the different days.
The cross section of flux tubes on each photospheric plane 
correspond to sunspots. 

Fig. 5.---
Examples of H$\alpha$ flares observed with DST at Hida Observatory.
Most of flares during 9 -- 12 March occurred at the east side and 
leading edge of F1.

Fig. 6.--- 
Flare positions and relative velocities of small umbrae to F1.
Almost all flares occurred around F1 from 9 through 12 March.
On March 9 and 11, many M-class and C-class flares occurred 
at the east side of F1.
On 15 March, before which the twisted magnetic flux bundle 
has already emerged out, no strong flare occurred around F1.

Fig. 7.---
H$\alpha \pm 0.0$ \AA $\;$ images obtained at (a) 02:31 
UT 11 March (b) 00:14 UT 12 March.
Notice the twisted dark loops at the east side of F1.
They have different orientations with each other, and they sometimes 
reconnect with each other to produce flares as shown in Fig. 5.

\pagebreak

\begin{center}
Table 1: Observational log

\begin{tabular}{rcc}\hline\hline
date & observation time (UT) & number of images \\ \hline
March 9 & 22:54 $\sim $ 07:29 & 11 \\
10 & 22:36 $\sim $ 06:58 & 5 \\
11 & 22:30 $\sim $ 02:01 & 7 \\
12 & 22:18 $\sim $ 04:39 & 6 \\
14 & 03:32 $\sim $ 06:33 & 2 \\
15 & 22:21 $\sim $ 05:57 & 7 \\ \hline
\end{tabular}
\end{center}

\end{document}